\documentclass[12pt]{article}

\usepackage{graphicx}

\title{Pion decay constants in  a strong magnetic field }
\author{  Yu.A.Simonov \\
State Research
Center\\Institute of Theoretical and Experimental Physics, \\
Moscow, 117218 Russia}

\newcommand{\beq}{\begin{eqnarray}}
 \newcommand{\eeq}{\end{eqnarray}}
\newcommand{\be}{\begin{equation}}
 \newcommand{\ee}{\end{equation}}

 \def\la{\mathrel{\mathpalette\fun <}}
\def\ga{\mathrel{\mathpalette\fun >}}
\def\fun#1#2{\lower3.6pt\vbox{\baselineskip0pt\lineskip.9pt
\ialign{$\mathsurround=0pt#1\hfil ##\hfil$\crcr#2\crcr\sim\crcr}}}

\newcommand{{\SD}}{\rm SD}

\newcommand{{\Mc}}{\mathcal{M}}

\newcommand{\ver}{\mbox{\boldmath${\rm r}$}}
\newcommand{\vesig}{\mbox{\boldmath${\rm \sigma}$}}

 \newcommand{\veA}{\mbox{\boldmath${\rm A}$}}

\newcommand{\veB}{\mbox{\boldmath${\rm B}$}}

\newcommand{\lan}{\langle}
\newcommand{\ran}{\rangle}

\begin{document}
\maketitle
\begin{abstract}

Decay  constants of the charged and neutral pions in magnetic field are
considered in the framework of the effective quark-antiquark lagrangian
respecting Gell-Mann--Oakes--Renner (GOR) relations at zero field. The
$\sqrt{\frac{e_qB}{\sigma}}$ dependence is found  in  strong fields $e_qB\gg
\sigma$ for  the neutral pion, while the charged pion constant decreases as
$\sqrt{\frac{\sigma}{e_qB}}$.

 \end{abstract}

 \section{Introduction}
 Pion decay constants are basic quantities in the chiral effective theory
 \cite{1,2} and  are present in the fundamental  GOR relations \cite{3}, moreover they play an
 important role of the order parameter, vanishing in the phase of restored
 chiral symmetry.

  The behavior of pion decay constants (pdc) in magnetic field (m.f.) allows to
  probe the most fundamental properties of the QCD vacuum and hadrons and
  therefore  together  with the behavior of chiral condensate was a hot topic
  in the  theoretical community, see \cite{4,5} for discussion and references.

   Specifically, in the case of pdc the analysis was done in the framework of
   the chiral perturbation theory (ChPT) in  \cite{6,7,8,8a}. It was argued in
   \cite{7}, that the parameter of the  ChPT is $  \xi = \frac{eB}{(4\pi
   f_\pi)^2}, \xi<1$ and it was found in \cite{6,7,8,8a}, that $f_\pi(eB)$ behaves
   in the lowest order as \be \frac{f^2_\pi(eB)}{f^2_\pi(0)} = 1 + \frac{2eB ln
   2}{(4\pi f_\pi (0))^2}+...\label{1}\ee

   In this analysis only pionic degrees of freedom contribute and the pion
   constituents, quarks, do not participate. At the same time it is clear, that
   quark should play an important role for $eB\ga \sigma$,  where $\sigma=0.18$
   GeV$^2$ is the string tension, and therefore the result (\ref{1}),
   neglecting the pion quark structure, should be modified for $eB\ga \sigma$,
   and possibly also for $eB\ga m^2_\pi$.

   Therefore it is of interest to study the pdc in m.f. in the approach of
   \cite{4,5}, where the explicit results were obtained for the  charged and
   neutral pion masses \cite{5} and chiral condensate \cite{4} as a function of
   m.f.

   It was found there, that in the case of $\pi^0$ the mass is strongly
   decreasing with $eB$ (in contrast to  much slower decrease in  \cite{6,7}),
   while for $\pi^+$ the mass is increasing (in agreement with lattice data
   \cite{9}). Moreover, in \cite{4} the chiral condensate was found to grow
   linearly with $eB$ in good quantitative agreement with lattice data
   \cite{10}, which contradict much smaller slope of ChPT \cite{7}.

   It is  therefore possible, that the results of ChPT are valid in a smaller
   region, moreover they need modification  for charged pions, since as was
   found in \cite{5}, GOR  relations are violated for charged pions in m.f.,
   while they are valid for neutral pions, in agreement with \cite{6}, \cite{7}.

   It is a purpose of the present paper to proceed in  the framework of the
   quark-antiquark formalism of \cite{4,5} to find the m.f. dependence of pdc
   both for neutral and charged pions. In the next section  the general
   formalism is shortly discussed and the resulting expression for pdc is
   given. Section 3 is devoted to the  neutral pions and section 4 to the
   charged pions, while section 5 contains summary and prospectives.

   \section{General formalism}

   The effective chiral quark-antiquark Lagrangian was derived and studied in
   \cite{11, 12,13,14} without m.f.,
   \be L_{ECL} =N_c tr log [(\hat \partial +m_f) \hat 1 + M\hat U],\label{2}\ee
   where the octet of Nambu-Goldstone (NG) mesons are given by
   \be \hat U =\exp (i\hat \phi \gamma_5), ~~\hat \phi = \phi_at_a, ~~
   a=1,...8.\label{3}\ee
   Here $M=M (x,y)$ is the  scalar confinement interaction defined via
   the vacuum average of field correlators $\lan tr F_{\mu\nu} (x) \phi F_{\mu\nu}
   (y) \phi\ran$, $ \phi(x,y) = P\exp ig \int^x_y  A_\mu dz_\mu$.

Expansion of (\ref{2}) to the  quadratic in $\hat\phi$ terms, produces the GOR
relations  and one obtains definitions of quark condensate $\lan \bar q q\ran$
and pdc \cite{11,12}, e.g. \be f^2_\pi = N_c M(0) \sum^\infty_{n=0}
\frac{|\psi_n (0)|^2}{m^3_n}, \label{4}\ee where \cite{12,13} \be  M(0) =
\frac{2 \sigma \lambda}{\sqrt{\pi}} (1+O(\sigma \lambda^2)) \approx 0.15~ {\rm
GeV}\label{5}\ee and $M_n, \psi_n$ are eigenvalues and eigenfunctions of the
$q\bar q$ Hamiltonian without chiral degrees of freedom. The corresponding
masses have been computed in \cite{11,12, 13} \be m_0 =0.4 ~{\rm GeV}, ~~ m_1
=1.35~{\rm GeV},~~m_2 =1.85 ~{\rm GeV}\label{6}\ee and resulting value of
$f_\pi$ is \cite{11,12}. \be f_\pi = 96 ~ {\rm MeV} \frac{M(0)}{(150~{\rm
MeV})}, ~- \frac{\lan \bar q q\ran}{n_f} =(217~{\rm MeV})^3.\label{9}\ee

Now one can include m.f as in \cite{4,5}, which is done using $\hat \partial
\to \hat D = \hat \partial - ie_q A_\mu^{(e)} \gamma_\mu$, where $\veA^{(e)} =
\frac12 (\veB \times \ver)$.

As  a result one obtains the following form of pdc \be f^2_\pi = N_c M^2 (0)
\frac12 \sum^\infty_{n=0} \left( \frac{|\psi^{(+-)}_{n,i}
(0)|^2}{(M^{(+-)}_{n,i})^3}+\frac{|\psi^{(-+)}_{n,i}
(0)|^2}{(M^{(-+)}_{n,i})^3}\right).\label{8}\ee

Here $(+-)$ and $(-+)$ refer to the spin projections of the quark and antiquark
respectively, and $n,i$ denote the  quantum numbers $n_\bot, n_3$  and  $u,d$
of the $q\bar q$ motion in m.f.

\section{The case of the neutral pion}

We now turn to the $q\bar q$ Hamiltonian in m.f. defining $M_{n,i}$ and
$\psi_{n,i}$ in (\ref{8}). In  the case of the neutral pion the corresponding
expression  without chiral degrees of freedom was derived before in
\cite{15,16} and studied for the case of pion in \cite{5},

\be M_n = \bar M_n^{(0)} + \Delta M_{\rm coul} + \Delta M_{SE} + \Delta
M_{ss}.\label{9}\ee

The form of $\bar M_n$ (prior to stationary point insertions $\omega_i \to
\omega_i^{(0)} (eB)$) is
 \be \bar M_n = \varepsilon_{n_\bot , n_z} +
\frac{m_1^2+\omega^2_1 - e_q\veB \vesig_1}{2\omega_1} +\frac{m_2^2+\omega^2_2 +
e_q\veB \vesig_2}{2\omega_2},\label{10}\ee where \be \varepsilon_{n_\bot, n_z}
= \frac{1}{2\tilde \omega} \left[ \sqrt{ e^2 B^2 + \frac{4\sigma\tilde
\omega}{\gamma}} (2n_\bot +1)+ \sqrt{\frac{4\sigma \tilde
\omega}{\gamma}}\left(n_z + \frac12\right)\right] + \frac{\gamma \sigma}{2}.
\label{11}\ee

We start  with  the $B=0$ case and write $\bar M_n$, Eq. (\ref{10}), for
$m_q=0, \omega_1=\omega_2$ , $e_q=e$.

\be \bar M_n = \omega + \frac32 \sqrt{ \frac{2\sigma \omega}{\gamma}} +
\frac{\gamma \sigma}{2}.\label{12}\ee Minimizing in $\omega, \gamma$ one
obtains \be \bar M_n (\omega_0, \gamma_0) = 4 \omega_0, ~~ \omega_0 =
\frac{\sqrt{3\sigma}}{2} = 0.367~ {\rm GeV},\label{13}\ee
 and $(\Delta M_{coul} + \Delta M_{se})$ cancel approximately 1/2 of $\bar M_n $
 ([4,5]), so that the final value of the mass in (\ref{8}) (without $(\Delta
 M_{ss}$) for $B=0$ is $m_0^{(+-)} = m_0^{(-+)}=2\omega_0$ Consider now the
 case of small $eB \ll \sigma$. In this case to the  lowest order in
 $\frac{eB}{\sigma}$ one obtains
 \be \frac{ |\psi^{(+-)} (0)|^2}{(M^{(+-)})^3}+ \frac{ |\psi^{(-+)}
 (0)|^2}{(M^{(-+)})^3}=2\left( \frac{\sigma}{2\pi}\right)^{3/2}
 \frac{1}{(2\omega_0)^3} +
 O\left(\left(\frac{eB}{\sigma}\right)^2\right).\label{14}\ee

 For $|\psi (0)|^2$ one has
 \be |\psi(0)|^2 = \frac{1}{\pi^{3/2} r^2_\bot r_3}, ~~ \frac{1}{r^2_\bot}
 =\frac12 \sqrt{(eB)^2 + \sigma^2 c}, ~~ \frac{1}{r_3} = \left( \frac{\sigma^2
 c}{4}\right)^{1/4},\label{15}\ee
  where $c=\frac{4\tilde \omega}{\gamma \sigma}$, and for $eB=0, ~ C(eB=0)=1$.

  At large $eB$, $eB\gg \sigma$ from (\ref{9}-\ref{11}) one obtains
  \be M_0^{(+-)}= 2 \omega_0^{(+-)}= 3^{-1/4}\sqrt{\sigma}, ~~ M_0^{(-+)} = 2
  \sqrt{2 eB}\label{16}\ee
   and for  $|\phi|^2$ one has from (\ref{15})
   \be \left| \psi^{(+-)}_{n_\bot=0, n_3}(0)\right|^2\cong \frac{\sqrt{\sigma}
   \sqrt{ e^2_q B^2 + \sigma^2}}{(2\pi)^{3/2}}\label{17}\ee

   \be \left| \psi^{(-+)}_{n_\bot=0, n_3}(0)\right|^2= \left(\frac{ {\sigma}}
   { 2\pi }\right)^{3/2} (c_{-+})^{3/4} \sqrt{1+ \left( \frac{e_qB}{\sigma} \right)^2 \frac{1}{c_{-+}}},
   \label{18}\ee where $ c_{-+} (B) =
   \left(1+\frac{8e_qB}{\sigma}\right)^{2/3} $ (cf \cite{4,5}).

   As a result one obtains for $eB\gg\sigma$ ( restoring $e=|e_q| \equiv e_q$)
   \be f^2_{\pi^0} (e_qB) \cong N_c M^2 (0) \frac{3^{3/4}}{2} \left(
   \frac{1}{2\pi}\right)^{3/2} \frac{e_qB}{\sigma} \cong f^2_{\pi^0} (0)  \frac{3^{3/4}}{2}\cdot 3^{3/2}
    \frac{e_qB}{\sigma} \cong \frac{5.9 e_qB}{\sigma}
    f^2_{\pi^0}(0),\label{19}\ee
 and finally, since $\pi^0=\frac{1}{\sqrt{2}} (| \bar u u\ran  +  | dd\ran )$,
 one obtains \be \overline{ f^2_{\pi^0} (eB)} =\frac12 \left( f^2_{\pi^0}
 \left( \frac23 eB\right) +  f^2_{\pi^0} \left( \frac{eB}{3}\right)
 \right).\label{20n}\ee

 The behavior of $f_{\pi^0} (eB)$ according to (\ref{20n}) is shown in Fig.1 in
 comparison to the ChPT result of \cite{6,7,8}.

 \section{The case of charged pions}

 In this case one can use the so-called  factorization approach, valid at large
 $eB\gg \sigma$, as shown in \cite{5}.  It is clear, that the  states  $ \rho^+
 (S_z=0)$ and  $\pi^+$ are mixed by the hyperfine interaction and tend to their
  asymptotic states $(+-)$ and $(-+)$ of  $(u\bar d)$ contribution. Considering
  $u$ and $\bar d$ states as independent, one has as in \cite{5}

\be M_{+-}(B) = \left (\sqrt{m^2_u+p^2_z}+ \sqrt{m^2_d+ p^2_z+
2|e_d|B}\right)_{P_z =0}\approx \sqrt{\frac23 eB}\label{20}\ee

\be M_{-+}(B) = \left(\sqrt{m^2_u+p^2_z+2e_u B}+ \sqrt{m^2_d+
p^2_z}\right)_{P_z =0}\approx \sqrt{\frac43 eB}.\label{21}\ee

Taking into account the hyperfine interaction, \cite{5}, one obtains the
asymptotic behavior of $\pi^+$ energy in m.f.

\be
 m_{as} (\pi^+ ) = M_{+-} (B) \approx \sqrt{\frac23 eB}.\label{22}\ee

 One can also deduce  from Eq. (29) of \cite{5}, that for the charges $e_u
 =\frac23 e $ and $e_{\bar d} = \frac{e}{3}$, the resulting  term in the
 hamiltonian const $B^2\eta^2_\bot$ is equal to $\frac{13 e^2 B^2}{18\cdot 32
 \omega} \eta^2_\bot$, which implies that $e^2_q $ for $\pi^+$ in  (\ref{17}),
 (\ref{18})
 should be replaced by approximately $\frac{13}{36} e^2$, $e_q =\sqrt{\frac{13}{36} } e$.

 Correspondingly, the $\pi^+$ decay constant is given by Eq. (\ref{8}), where
 masses  can be taken from (\ref{22}) and $|\psi (0)|^2$ from  (\ref{17}), (\ref{18}).
 Results will be discussed in the next section.

 \section{Discussion and results}

 We start with the case of $\pi^0$ meson, where $f^2_{\pi^0}$ is given in
 (\ref{8}), (\ref{20}), $n = n_\bot, n_3$ and $i=u,d$, and $ M^{(-+)}_{n,i}$
 are nonchiral pion masses, given by the eigenvalues of the Hamiltonian in Eqs.
 (\ref{9}), (\ref{10}), (\ref{11}). As is was found above (see also \cite{5})
 the mass $M_{oi}^{(+-)}$ changes from approximately  $\sqrt{ 3\sigma}- \Delta
 M_{coul} - \Delta M_{SE}- \Delta M_{ss}\cong \sqrt{\sigma}$ at $B=0$ to
 $3^{-3/4} \sqrt{\sigma}$ at $eB\gg \sigma$, which can be approximated by the
 relation
 \be M^{(+-)}_{o,i} = \sqrt{\sigma} \left( \frac{ 1+ \left( \frac{e_i
 B}{\sigma}\right)^2}{\left( \frac{e_i B}{\sigma}\right)^2 3^{3/2} +1
 }\right)^{1/2}\equiv (M_{o,i}^{(+-)} (0)  \mu^{(+-)}
 \left(\frac{e_iB}{\sigma}\right), ~~ i=u,d,\label{5.1}\ee
 with $\mu^{(+-)} (x) =1 $ at $x=0$ and $3^{-3/4}$ at $x=\infty$.

 For the $|\psi(0)|^2$ one has from (\ref{15})

\be |\psi^{(+-)}_{o,i} (0)|^2=\left( \frac{\sigma}{2\pi}\right)^{3/2}\sqrt{
c_{+-} + \left(\frac{e_iB}{\sigma}\right)^2}, ~~ i=u,d,\label{5.2}\ee
 and $c_{+-}(eB)$ changes  from  1 for $eB=0$ to $3^{-1/12}\cong 0.91$ at $eB\gg
 \sigma$, and we take $c_{+-} =1$ with this accuracy for all $eB$.

 For the $(-+)$ states one has

 \be M^{(-+)}_{o,i} = M^{(-+)}_{o,i} (0) \mu^{(-+)} \left( \frac{e_i
 B}{\sigma} \right) , ~~  \mu^{(-+)} (x) \cong \left(  { 2^3 x +
 1} \right)^{1/2}.\label{5.3}\ee

 Finally, for $\psi^{(-+)} (0)$ one obtains from (\ref{15}) (see also \cite{5}.
 Eq.(71))

 \be   |\psi^{(-+)}_{o,i} (0)|^2= (c_{-+})^{1/4} \left( \frac{\sigma}{2\pi}\right)^{3/2}\sqrt{
c_{-+} + \left(\frac{e_iB}{\sigma}\right)^2}, ~~ c_{-+} = \left( 1+ \frac{8e_i
B}{\sigma}\right)^{2/3}.\label{5.4}\ee

As a result one obtains for $f_{\pi^0}$ (approximating the sum in (8) by the
leading first term) \be \frac{f^2_{\pi^0} (eB) }{ f^2_{\pi^0} (0)}=\frac14
\left(
 \sum_{i=u,d} \frac{ \sqrt{ 1+ x^2_i}}{(\mu^{+-} (x_i))^3}+
\sum_{i=u,d} \frac{  c^{1/4}_{-+} (x_i) \sqrt{  c_{-+} (x_i) + x^2_i}}
 {(\mu^{-+ } (x_i))^3}\right)\label{5.5}\ee
 where $x_i = \frac{|e_i |B}{\sigma}$. One can see, that at large $eB\gg
 \sigma$ the behavior is
 \be \frac{f^2_{\pi^0} (eB)}{ f^2_{\pi^0} (0)} \cong    2.96 \frac{eB}{\sigma} ,
  ~~ e = e_u + | e_d|, \label{5.6}\ee

 \begin{figure}[h]
  \centering
  \includegraphics[width=9cm]{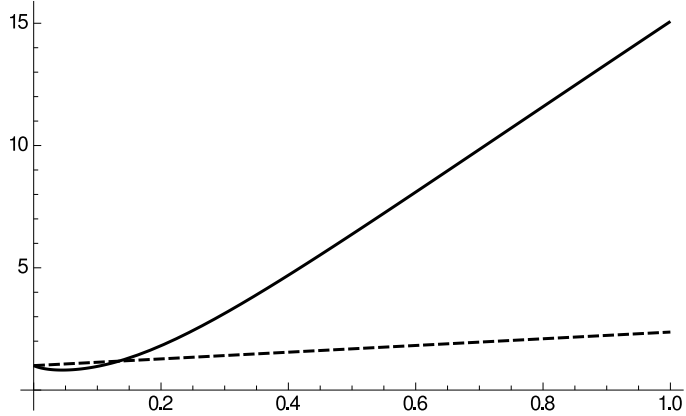}
 \caption{ The ratio $\frac{ f^2_{pi^0} (eB)}{f^2_{\pi^0} (0)}$ as a function
 of  $ x= \frac{eB}{1~{\rm GeV}^2}$  in the ChPT (the lower curve
  (dashed)) and according to Eq.(28) (the upper curve)}
\end{figure}

The behavior of the ratio (\ref{5.5}) is depicted in Fig. 1 together with the
prediction of the ChPT \cite{6,7}. One can see, that the ratio (\ref{5.5})
grows much faster and exceeds the ChPT prediction more than 7 times at $eB =1$
GeV$^2$. However at small $eB$, $ eB\ll \sigma$ the chiral ratio (\ref{1})
grows $1+ d_{ch} \left( \frac{eB}{\sigma}\right)$, while the $q\bar q$ answer
(\ref{5.6}) is $1+ d_{q\bar q} \left( \frac{eB}{\sigma}\right)^2$, and  $
d_{ch} =\frac{ 2ln 2\sigma^2}{(4\pi f_\pi  (0))^2} \approx 0.04, ~~ d_{q\bar q}
\approx O(1)$.

 Hence there appears a region  $f^2_\pi (0) \sim eB \ll \sigma$, where the
 result of ChPT is dominant, and the $q\bar q$ structure of $\pi^0$ is not yet
 displayed.

 It is also interesting to follow the fate of the GOR relations. Writing it in
 the form, averaged over flavors
 \be m^2_{\pi^0}f^2_{\pi^0}= \frac{m_n+m_d}{2} |\lan\overline{\bar q q}\ran|,~~
|\lan\overline{\bar q q}\ran|= \frac12 \sum_{i=u,d} (\bar q q_i)\label{5.2}\ee
where $\lan \bar q q \ran$ is \cite{4,5}

\be |\lan \bar q q\ran_i (B) | =| \lan \bar q q\ran_i(0) | \frac12 \left\{
\sqrt{1+\left(\frac{e_qB}{\sigma}\right)^2}+
\sqrt{1+\left(\frac{e_qB}{\sigma}\right)^2\frac{1}{c_{-+}}}\right \}
\label{5.8a}\ee

Comparing (\ref{5.2}) and (\ref{8}), one can see, that in(\ref{5.2}) both
l.h.s. and r.h.s. have the same form of the numerator, $|\psi (0)|^2$, and
differ only in the power of $M^{(+-)}_{n,i}$ in the denominator, and since
$M_{\pi^0} (eB) $ is  proportional to $M_{o,i}^{(+-)},$
 the GOR relation for
$\pi^0$ holds also large $eB$, in agreement with conclusions of \cite{4,5}.

We now turn to the case of the charged pion. In this case at large $eB$ one can
use factorization technic \cite{5}, assuming both $u$ and $\bar d$ quark
independent of each other, since at $eB\gg \sigma$ the main  interaction term
$\lan \sigma |\ver_1 - \ver_2|\ran$ is subleading as compared to the m.f.
contribution $\sqrt{ |e_iB|}.$

As it was found in \cite{5}, the asymptotic form of the energy \be M_{\pi^+}
(eB) \approx M^{(+-)} (B) \approx \sqrt{\frac23 eB}. \label{5.8}\ee At this
point it is interesting to compare this result with an  exact solution, which
obtains in the case  $ e_u= e_{\bar d} = \frac{e}{2}$, see \cite{16} for
details. In this case the mass can be written as (the $ \left (\frac{e}{2},
\frac{e}{2}\right)$ approximation) \be  \frac{ M_{\frac{e}{2}, \frac{e}{2}}
(eB)}{M_{\frac{e}{2}, \frac{e}{2}} (0)} = \sqrt{1+\chi (eB)
\frac{eB}{\sigma}},\label{33}\ee where $\chi(eB) \approx 1.22 \left( \frac{1 +
eB/\sigma}{1+2eB/\sigma}\right).$ One can see, that at large
$\frac{eB}{\sigma}$ the ratio of (\ref{33}) and (\ref{5.8}) is 1.04, so that
one can  use  the $\left( \frac{e}{2}, \frac{e}{2}\right)$ approximation to
calculate $|\psi (0)|^2$,   which otherwise  is difficult to do  in the
factorization
 scheme.
 In this case,  using Hamiltonian (72) of \cite{16}, one obtains approximately

 \be \frac{ |\psi_{\frac{e}{2},
\frac{e}{2}}(0)|^2_{ eB}}{|\psi_{\frac{e}{2}, \frac{e}{2}}(0) |^2_0} \cong
\left( 1+ \frac{1}{16} \chi^2 (eB) \frac{eB}{\sigma}\right)^{1/2}.\label{34}\ee
As a result the behavior of the $f_{\pi^+}$ can be written as

 \be \frac{f^2_{\pi^+} (eB)}{f^2_{\pi^+} (0)}=
 \frac{
\left( 1+ \frac{1}{16} \chi^2 (eB)\left(
\frac{eB}{\sigma}\right)^2\right)^{1/2}}{ \left( 1+ \chi  (eB)
\frac{eB}{\sigma} \right)^{3/2}}.\label{35}\ee
 From  (\ref{35}) one can deduce, that for $eB\la 1$ GeV$^2$ the ratio
 (\ref{35}) behaves as $\frac{1}{\left( 3.4 \frac{eB}{1{\rm ~
 GeV}^2}\right)^{3/2}}$, which transforms into a slower decrease, $
 \frac{1}{\sqrt{eB}}$ at large $eB$.

 \begin{figure}[h]
  \centering
  \includegraphics[width=9cm]{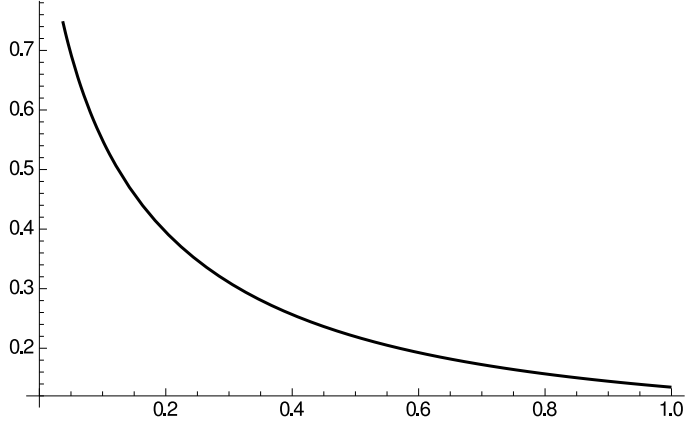}
 \caption{  The ratio $\frac{ f^2_{pi^+} (eB)}{f^2_{\pi^+} (0)}$ as a function
 of  $ x= \frac{eB}{1~{\rm GeV}^2}$}
\end{figure}

In Fig. 2 we show \footnote{We have excluded the region $eB\ll \sigma$, where
our result (\ref{35}) is less accurate.} the ratio (\ref{35}) as a function of
$eB$ up to $eB = 1$ GeV$^2$. One can see a strong decrease of $f^2_{\pi^+}(eB)$
with growing $eB$. At the same time  the $\pi^+$ mass is growing as
$\sqrt{\tilde e B}$,  $\tilde e =\frac23 e$ \cite{5} (a similar behavior is
found on  the lattice \cite{9} with $\frac23   e \leq\tilde e \leq e$, so that
the l.h.s. of the GOR relation is kept constant, while in  the r.h.s. the quark
condensate $\lan \bar q q \ran |$ is growing as $eB$ \cite{4}. This is a clear
manifestation of the fact, that GOR relations are violated for charged pions in
m.f. -- a conclusion, which was made  before in ChPT \cite{6,7,8}.

Summarizing, we  have found the  m.f. dependence of the decay constants for the
neutral and charged pions. We have compared the neutral pion constant behavior
with that obtained in the ChPT,  and found strong disagreement at large m.f.
$eB\gg\sigma$, while for small m.f., $eB \la f^2_{\pi^0}$, the ChPT prevails.
 A
moderate increase of $\frac{ f^2_{\pi^0}(eB)}{f^2_{\pi^0}(0)}$, similar to the
ChPT prediction, was also found in the NJL model in \cite{18}.

We have also found the  decreasing behavior of $f_{\pi^+} \sim (eB)^{-1/2}$ for
large $eB$ and confirmed the violation of GOR relations for charged pions in
m.f.

The author is grateful to N.O.Agasian for useful discussions and suggestions
and to M.A.Andreichikov for discussions and helpful assistance. The financial
support of the RFBR grant 1402-00395 is acknowledged.


\begin{thebibliography}{99}
%

\bibitem{1}
  S.~Weinberg, Physica {\bf A 96}, 327 (1979).


\bibitem{2} J.~Gasser and H.~Leutwyler, Ann. Phys. (NY) {\bf 158}, 142 (1984);
Nucl. Phys. {\bf B250}, 465 (1985).

\bibitem{3}
M.~Gell-Mann, R.L~Oakes, and B.~Renner, Phys. Rev. {\bf 175}, 2195 (1968).


\bibitem{4}  Yu.~A.~Simonov, JHEP  {\bf 1401},  118 (2014) arXiv:
1212.3118 [hep-ph].


\bibitem{5}  Yu.~A.~Simonov, JHEP  {\bf 1309},  135 (2013) arXiv:
1306.2232 [hep-ph].

\bibitem{6} I.~A.~Shushpanov and A.~V.~Smilga,  Phys. Lett. {\bf B402}, 351
(1997), [hep-ph/9703201].



\bibitem{7} N.~O.~Agasian and I.~A.~Shushpanov,  JETP Lett.\  {\bf 70}, 717 (1999);
Phys.\ Lett.\ B {\bf 472}, 143 (2000); JHEP {\bf 0110}, 006 (2001).

\bibitem{8}  N.~O.~Agasian,   Phys.\ Lett.\ B {\bf 488}, 39 (2000); Phys.\ Atom.\ Nucl.\  {\bf 64}, 554 (2001).

\bibitem{8a}   J.~O.~Andersen, JHEP {\bf 1210}, 005 (2012);  Phys.\ Rev.\ D {\bf 86}, 025020 (2012).

\bibitem{9} G.S.Bali, F.Bruckmann, G.Endr\"{o}di et al., JHEP \textbf{1202}, 044 (2012).

\bibitem{10} G.S.Bali, F.Bruckmann, G.Enr\"{o}di, Z.Fodor, S.D.Katz and
A.Sch\"{a}fer, Phys. Rev. D {\bf 86},071502 (2012); arXiv:1206.4205; M.D'Elia,
Lect. Notes Phys. {\bf 871}, 181 (2013).

\bibitem{11}  Yu.~A.~Simonov,
Phys. Rev. D {\bf  65}, 094018 (2002); hep-ph/0201170.


\bibitem{12}  Yu.~A.~Simonov,
 Phys. Atom. Nucl. {\bf 67}, 846 (2004); hep-ph/0302090.


\bibitem{13} Yu.~A.~Simonov,  Phys. Atom. Nucl. {\bf 67}, 1027 (2004);
hep-ph/0305281.


\bibitem{14} S.~M.~Fedorov and
Yu.~A.~Simonov,  JETP Lett. {\bf 78}, 57 (2003); hep-ph/0306216.

\bibitem{15}
 M.~A.~Andreichikov, B.~O.~Kerbikov, V.~D.~Orlovsky and
Yu.~A.~Simonov,  Phys. Rev. {\bf  D 87}, 094029 (2013); arXiv:1304.2533
[hep-ph].

\bibitem{16}  Yu.~A.~Simonov,
Phys. Rev.  {\bf D 88}, 025028 (2013); arXiv:1303.4952 [hep-ph].

\bibitem{18} Sh.Fayazbakhsh and N.Sadooghi, Phys. Rev.  D {\bf 88}, 065030
(2013); arXiv:1306.2098; Sh.Fayazbakhsh, S.Sadeghian,  and  N.Sadooghi, Phys.
Rev. D {\bf 86}, 085042 (2012); arXiv:1206.6051.

\end{thebibliography}
   \end{document}